\documentclass[amssymb,12pt,showpacs,aps]{revtex4}
\usepackage{graphicx}
\begin{document}
\title{Periodic orbit theory and spectral statistics 
for scaling quantum graphs}
\date{\today}
\author{Yu. Dabaghian}
\affiliation{Department of Physiology,
Keck Center for Integrative Neuroscience,\\
University of California,
San Francisco, California 94143-0444, USA
E-mail: yura@phy.ucsf.edu}

\begin{abstract}
The explicit solution to the spectral problem of quantum 
graphs found recently in \cite{Anima}, is used to produce 
the exact periodic orbit theory description for the probability 
distributions of spectral statistics, including the distribution 
for the nearest neighbor separations, $s_{n}=k_{n}-k_{n-1}$, and 
the distribution of the spectral oscillations around the average, 
$\delta k_{n}=k_{n}-\bar k_{n}$.
\end{abstract}

\pacs{05.45.+b,03.65.Sq,72.15.Rn}
\maketitle

\section{Introduction}

Quantum graphs consist of a quantum particle moving on a quasi 
one-dimensional network. In the limit $\hbar=0$, quantum networks 
produce a nonintegrable classical counterpart - a classical particle 
moving on the same network, scattering randomly on its vertexes 
\cite{QGT1,QGT2}. As shown in \cite{Gaspard1}, this stochastic dynamics 
is characterized by an exponential proliferation of periodic orbits, 
positive Kolmogorov entropy and other familiar features of finite 
dimensional deterministic chaotic systems.

Such classical nonintegrability is clearly manifested in quantum 
regime. Quantum networks provide excellent illustrations to many 
general concepts, phenomenological hypotheses and mathematical 
constructions of quantum chaology. For example, extensive numerical 
studies \cite{QGT2,Severini,GnutzSeif} and analytical \cite{GnutzAtl1,
GnutzAtl1,Berkolaiko2,Berkolaiko3,Bogomolny1,Tanner,Dahlqvist,Berry1} 
have demonstrated that the statistics of the nearest neighbor spacings 
distribution, the two point autocorrelation function, the form factor, 
and the spectral rigidity of the quantum graph spectra are close to the 
ones predicted by the random matrix theory (RMT). Since the latter three 
statistics can be expressed in terms of the spectral density functional, 
they were also studied analytically in terms of the Gutzwiller's periodic 
orbit series expansion \cite{Severini,GnutzSeif,Berkolaiko1,Berkolaiko2,
Berkolaiko3,Gnutzmann,Tanner,Bogomolny1,Dahlqvist,Berry1,Berry2}.

Due to the relatively simple structure of the network dynamics, the 
results of the periodic orbit theory analysis of quantum graphs are 
particularly complete. Moreover, the periodic orbit expansions which 
usually have semiclassical accuracy, are exact for the quantum networks 
and can be viewed as mathematical theorems. In addition, as it was shown 
recently in \cite{Opus,Prima,Anima}, the periodic orbit theory for the 
quantum graphs can describe not only the global characteristics of the 
spectrum (e.g. the density of states, spectral staircase, quantum and 
classical dynamical zeta functions, etc.), but also the \emph{individual} 
eigenvalues of the energy or the momentum.

This fact provides an interesting opportunity to study several additional 
statistical distributions, including those that are not directly accessible 
via the Gutzwiller's expansion for the density of states, such the distribution 
of the eigenvalue fluctuations around the average, $\delta_n=k_{n}-\bar k_{n}$, 
the nearest neighbor spacings, $s_n=k_n-k_{n-1}$, etc., which is the main 
subject of this paper.

The paper is organized as following: Section II reviews the spectral 
hierarchy method \cite{Anima}. Section III discusses the statistical 
spectral distributions for the regular quantum graphs, which are later 
generalized for the irregular graphs in Sections IV and V. A short 
discussion of certain statistical universality aspects of the resulting 
distributions is given in the Section VI.

\section{Spectral hierarchy for quantum networks}

The idea of producing the individual momentum eigenvalues 
$k_{n}\equiv \hat k^{(0)}_{n}$ is based on using the periodic orbit 
expansion for the density of states, 
\begin{equation}
\rho^{(0)}(k)\equiv\sum_{n=1}^{\infty}\delta\left(k-\hat k^{(0)}_{n}\right),
\label{rhopeaks}
\end{equation}
and an auxiliary sequence $\hat{k}_{n}^{(1)}$, that separates the spectral 
points 
from one another: 
\begin{equation}
\hat{k}_{n-1}^{(1)}<\hat k^{(0)}_{n}<\hat{k}_{n}^{(1)},\ \ \ n=1,...\,\,.
\label{separators}
\end{equation}
From these two constituents one can obtain the quantum energy levels via 
\begin{equation}
\hat k^{(0)}_{n}=
\int_{\hat{k}_{n-1}^{(1)}}^{\hat{k}_{n}^{(1)}}\rho^{(0)}(k)\,kdk.
\label{regularroots}
\end{equation}
If any sequence with property (\ref{separators}) is known as a global 
function of $n$, $\hat{k}_{n}^{(1)}=\hat{k}^{(1)}(n)$, the relationship 
(\ref{separators}) produces the explicit solution to the spectral problem, 
\begin{equation}
k_n=\hat k^{(0)}_{n}=\hat k^{(0)}(n).
\label{knformula}
\end{equation}
The explicit integration in (\ref{regularroots}) is possible due to the 
exact periodic orbit expansion for the density of states. As shown in 
\cite{QGT1,QGT2,Nova}, the exact expansion for $\rho^{(0)}(k)$, has the 
form 
\begin{equation}
\rho^{(0)}(k)=\frac{L_{0}}{\pi}+\frac{1}{\pi}\mathop{\rm Re}
\sum_{p}L_{p}^{(0)}A_{p}^{(0)}e^{iL_{p}^{(0)}k}.
\label{rhoseries}
\end{equation}
where $L_{p}^{(0)}$ and $A_{p}^{(0)}$ are correspondingly the action 
length and the weight factor of the periodic orbit $p$, $L_{0}$ is the 
total action length of the network. For the scaling quantum graphs the 
weight coefficients, $A_{p}^{(0)}$, are $k$-independent.

\begin{figure}[tbp]
\begin{center}
\includegraphics{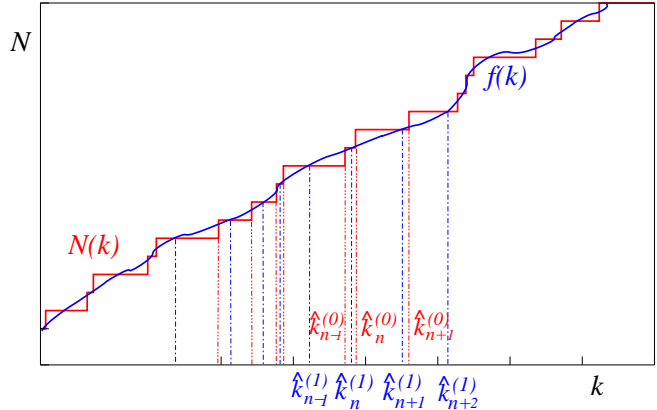}
\end{center}
\caption{A continuous function $f(k)$ that pierces every stair step 
of the spectral staircase $N(k)$ of an 4 vertex linear chain graph.
The intersection points $\hat{k}_{n}^{(1)}$ separate the momentum
eigenvalues $\hat{k}_{n}^{(0)}$.}
\label{Function1}
\end{figure}

Effectively, obtaining the spectral points as a function of their index 
is equivalent to ``inverting'' the spectral staircase function, 
\begin{equation}
N(k)=\sum_{i}\Theta (k-k_{i}),
\label{staircase}
\end{equation}
i.e. passing from $N(k_{n})=n$ to $k(n)=N^{-1}(n)$. Geometrically, finding 
an auxiliary sequence that singles out separate peaks in (\ref{rhopeaks}), 
amounts to finding a suitable monotone function $f(k)$, whose graph 
intersects every stair step of the spectral staircase. As it is illustrated 
on Fig.\ref{Function1}, the intersection points, $\hat k^{(1)}_{n}$,
\begin{eqnarray}
f\left(\hat{k}_{n}^{(1)}\right) =N\left(\hat{k}_{n}^{(1)}\right) =n,
\label{fk}
\end{eqnarray}
clearly satisfy the condition (\ref{separators}), so solving the equation 
(\ref{fk}) would yield the whole sequence $\hat{k}_{n}^{(1)}$ as a single 
globally defined function of the index $n$. 
%%%%%%%%%%%%%%%%%%%%%%%%%%%%%%%%%%%%%%%%%%%%%%%%%%%%%%%%%%%%%%%%%%%%%%
%%%%%%%%%%%%%%%%%%%%%%%%%%%%%%%%%%%%%%%%%%%%%%%%%%%%%%%%%%%%%%%%%%%%%%
It is well known however, that following the behavior of the spectral staircase 
function (\ref{staircase}) in such detail is generally a difficult task (see e.g. 
\cite{BaltesHilfe}). Luckily, quantum graphs allow an alternative approach, 
\cite{Opus,Prima,Anima}, which is based on certain properties of their spectral 
determinant.

As shown in \cite{QGT2,Opus,Prima,Anima}, the spectral determinant $\Delta(k)$ 
for quantum graphs is a finite order exponential sum, 
\begin{equation}
\Delta (k)=1+e^{i2(L_{0}k-\pi \gamma_{0})}-\sum_{i=1}^{N_{\Gamma}}
a_{i}e^{i2(L_{i}k-\pi\gamma_{i})},  \label{spectraldet}
\end{equation}
with constant $a_{i}$, $L_{i}<L_{0}$ and $\gamma_{i}$. The order $N_{\Gamma}$
of the sum depends on the topology of the graph $\Gamma$. The explicit form
(\ref{spectraldet}) can obtained from imposing the boundary conditions on the 
wave function of the quantum particle moving on the network, e.g. by using the 
scattering quantization \cite{QGT2,Opus} or the Bogomolny's transfer operator 
\cite{Bogomolny2} methods. The roots of $\Delta(k)$ define the quantum spectrum 
of the momentum, $\Delta(\hat k^{(0)}_{n})=0$.

\begin{figure}[tbp]
\begin{center}
\includegraphics{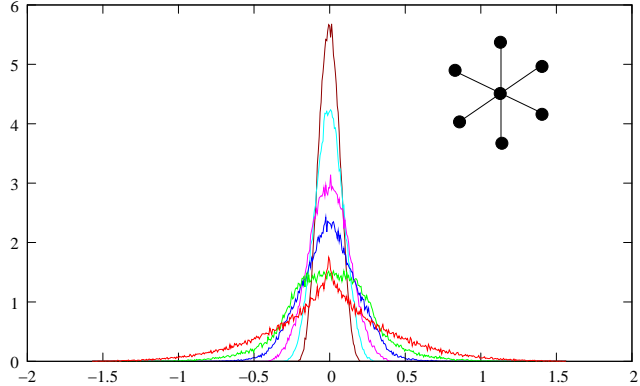}
\end{center}
\caption{Histogram of the fluctuations of the roots of the 6-star graph 
of irregularity degree 5 and of the following 5 separating sequences, 
$\delta_{n}^{(0)}$, ..., $\delta_{n}^{(5)}$. The initial spread of the 
fluctuations $\delta_{n}^{(0)}$ becomes progressively narrow for the 
higher values of $j$.}
\label{Fs_Quadrangle}
\end{figure}

There are three key properties of the spectral determinant $\Delta (k)$ relevant 
for the following discussion. First, its roots as well as the roots of all of its
derivatives are real \cite{Laguerre,Levin}. Second, there is exactly one root of 
its $j$th derivative, $\Delta^{(j)}(k)$, between every two neighboring roots of 
$\Delta^{(j+1)}(k)$. This implies that the zeroes of $\Delta(k)$ are interlaced 
by the zeroes of $\Delta'(k)$, as required by (\ref{separators}), which in turn 
are interlaced by the zeroes of $\Delta''(k)$ and so on \cite{Laguerre,Levin}. 
Lastly, as shown in \cite{Anima}, the higher is the order of the derivative, the 
more orderly is the behavior of the roots of $\Delta^{(j)}(k_{n}^{(j)})=0$ (Fig.
\ref{Fs_Quadrangle}). 
In fact, for any quantum graph system, there exists a finite integer $r$ (called 
the regularity degree of the graph in \cite{Anima}, see also \cite{Bogomolny3},) 
such that the roots of $\Delta^{(r)}(k)$ can be interlaced by a periodic sequence 
of points 
\begin{equation}
\hat{k}_{n}^{(r+1)}=\frac{\pi}{L_{0}}\left(n+\frac{1}{2}\right).
\label{persep}
\end{equation}
According to the criterion used in \cite{Opus,Prima,Anima}, 
the regularity degree $r$ is the minimal integer for which the inequality 
\begin{eqnarray}
\sum_{i}\left\vert a_{i}\left(\frac{L_{i}}{L_{0}}\right)^{r}\right\vert <1
\end{eqnarray}
holds. Since $\left(\frac{L_{i}}{L_{0}}\right)<1$, this criterion allows to find 
a finite regularity degree for every quantum graph system using the coefficients 
of the spectral determinant (\ref{spectraldet}).
%%%%%%%%%%%%%%%%%%%%%%%%%%%%%%%%%%%%%%%%%%%%%%%%%%%%%%%%%%%%%%%%%%%%%%
%%%%%%%%%%%%%%%%%%%%%%%%%%%%%%%%%%%%%%%%%%%%%%%%%%%%%%%%%%%%%%%%%%%%%%
Hence, there exist $r+1$ almost periodic sequences of interest, $\hat{k}_{n}^{(j)}$, 
$j=0$, ..., $r$, such that 
\begin{equation}
\Delta^{(j)}\left(\hat{k}_{n}^{(j)}\right) =0, 
\label{spectraleq}
\end{equation}
and 
\begin{equation}
\hat{k}_{n-1}^{(j)}<\hat{k}_{n}^{(j-1)}<\hat{k}_{n}^{(j)}.  
\label{bootstrap}
\end{equation}
The density functional for the $j$th sequence,
\begin{eqnarray}
\rho^{(j)}(k) =\sum_{n}\delta \left(k-\hat{k}_{n}^{(j)}\right) ,
\end{eqnarray}
allows to pinpoint the exact location of $\hat{k}_{n}^{(j-1)}$ on the interval between 
$\hat{k}_{n}^{(j)}$ and $\hat{k}_{n-1}^{(j)}$,
\begin{equation}
\hat{k}_{n}^{(j-1)}=\int_{\hat{k}_{n-1}^{(j)}}^{\hat{k}_{n}^{(j)}}
\rho^{(j-1)}(k)\,kdk,  
\label{seps}
\end{equation}
for $\quad j=r+1,r-1,...,1$ and $n=1$,... . Since the roots of each $\Delta^{(j)}(k)$ 
form an almost periodic set \cite{Levin} the density functional of each of the sequences 
$k^{(j)}_{n}$ can be expanded into an explicit harmonic series, which allows to evaluate 
the integral (\ref{seps}) explicitly for every $n$ and $j$.

The strategy that allows obtaining the separating sequences $\hat{k}_{n}^{(j)}$ and 
eventually getting the physical spectral sequence $\hat{k}_{n}^{(0)}$, follows directly 
from the bootstrapping (i.e. interlacing) property of the sequences (\ref{bootstrap}) and 
the relationship (\ref{seps}). One first finds the sequence $\hat{k}_{n}^{(r)}$ starting 
from the periodic separators (\ref{persep}), then uses it to find $\hat{k}_{n}^{(r-1)}$, 
and so on. After $r$ steps of moving up the hierarchy the spectrum $\hat{k}_{n}^{(0)}$ 
is produced \cite{Anima}.

\begin{figure}[tbp]
\begin{center}
\includegraphics{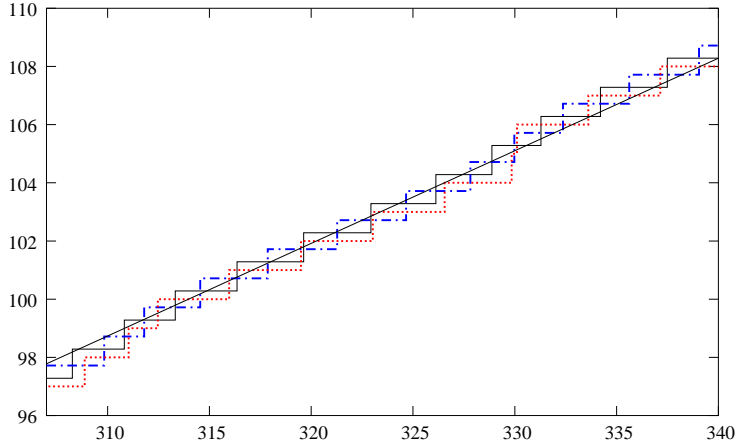}
\end{center}
\caption{The bootstrapping of the spectral staircase (dotted line) by the 
$N^{(1)}$ staircase (dash-dotted line), bootstrapped in turn by the $N^{(2)}$ 
staircase (solid line) intersected by the Weyl's average (straight line).
Note that the $N^{(2)}$ staircase does not bootstrap the $N^{(0)}$. These 
graphs were obtained for the dressed 4 vertex linear chain graph \cite{Anima}.}
\label{Bootstrapping}
\end{figure}

Geometrically, this algorithm can illustrated using the integrated densities 
\begin{eqnarray}
N^{(j)}(k) =\sum_{n}\Theta \left(k-\hat{k}_{n}^{(j)}\right).
\end{eqnarray}
The separating property of the $j$th sequence $\hat{k}_{n}^{(j)}$ with respect to the 
sequence $\hat{k}_{n}^{(j-1)}$ implies that the staircase $N^{(1)}(k)$ ``bootstraps'' 
the staircase $N^{(0)}(k)$, whereas $N^{(2)}(k)$ bootstraps $N^{(1)}(k)$ and so on (see 
Fig.\ref{Bootstrapping}, compare to Fig.\ref{Function1}).
%%%%%%%%%%%%%%%%%%%%%%%%%%%%%%%%%%%%%%%%%%%%%%%%%%%%%%%%%%%%%%%%%%%%%%
%%%%%%%%%%%%%%%%%%%%%%%%%%%%%%%%%%%%%%%%%%%%%%%%%%%%%%%%%%%%%%%%%%%%%%
The final staircase $N^{(r)}(k)$ is pierced by the Weyl's average 
\begin{equation}
\bar{N}(k) =\frac{L_{0}}{\pi}k-\frac{1}{2}.
\end{equation}
In the simplest case of $r=0$, the spectral points themselves can be separated from 
one another by a periodic sequence (\ref{persep}). Geometrically, this implies that 
the Weyl's average pierces every stair step of the original $N(k)$, which guarantees 
the existence of periodic separators (\ref{persep}). Such networks were referred to 
as the ``regular graphs'' in \cite{Opus,Prima,Anima}. The result of integration 
(\ref{regularroots}) in this case produces the quantum eigenvalues in the form of the 
periodic orbit series, 
\begin{equation}
k_{n}=\frac{\pi}{L_{0}}n-\frac{2}{L_{0}}\sum_{p}\frac{A_{p}^{(0)}}{\omega^{(0)}_{p}}
\sin \left(\frac{\omega^{(0)}_{p}}{2}\right) \sin \left(\omega^{(0)}_{p}n\right),
\label{kn}
\end{equation}
where $\omega^{(0)}_{p}=\pi L_{p}^{(0)}/L_{0}$. The first term in (\ref{kn}) gives the average 
behavior of the eigenvalue sequence, $\bar k_{n}=\frac{\pi n}{L_{0}}$, and the subsequent 
periodic orbit sum describes zero-mean fluctuations of $k_{n}$s around the average. It 
will be more convenient to describe the fluctuations in terms of the quantity 
$\delta_{n}^{(0)}=L_{0}\left(k-\bar{k}_{n}\right)/\pi$, $n=1,2,...$, 
\begin{equation}
\delta^{(0)}_{n}=-\frac{2}{\pi}\sum_{p}\frac{A_{p}^{(0)}}{\omega^{(0)}_{p}}
\sin\left(\frac{\omega^{(0)}_{p}}{2}\right) \sin \left(\omega^{(0)}_{p}n\right).
\label{deltakn}
\end{equation}
As demonstrated in \cite{Opus,Prima,Anima}, if the
sum (\ref{kn}) includes only the orbits that involve a certain fixed number 
$m$ of scatterings at the vertexes of the graph, it produces the $m$th order
approximation to the exact value of $k_{n}$ for each $n$.

\section{Eigenvalue distribution for regular graphs}

Let us first study the spectral fluctuation statistics for the regular
graphs based on the expansion (\ref{kn}). The transition to statistical
description of the sequence $\delta_{n}^{(0)}$, $n=1,2,...$, can be made
based on the properties of the sequence of the remainders 
\begin{equation}
x_{n}=\left[ \alpha n\right]_{\mathop{\rm mod}1},\ n=1,2,...,  \label{xn}
\end{equation}
It is a well known number-theoretic result \cite{Karatsuba,Kuipers} that 
the sequence (\ref{xn}) is uniformly distributed over the interval 
$x_{n}\in\left[0,1\right]$ for any irrational number $\alpha$.

For every periodic orbit $p$ in (\ref{deltakn}), the frequency $\omega^{(0)}_{p}$
is defined as 
\begin{equation}
\omega^{(0)}_{p}=m^{(0)}_{p,1}\Omega_{1}+...+m^{(0)}_{p,N_{B}}\Omega_{N_{B}}=
\left\langle \vec m^{(0)}_{p},\vec{\Omega}_{p}\right\rangle ,
\label{omegas}
\end{equation}
where $\Omega_{i}=\pi l_{i}/L_{0}$ are the ``bond frequencies'' defined by 
the bond lengths $l_{i}$, $\sum_{i}\Omega_{i}=1$, and the vector 
$\vec m^{(0)}_{p}=(m^{(0)}_{p,1},...,m^{(0)}_{p,N_{B}})$ gives the number 
of times the orbit $p$ passes over the bond $i$. Considering that for every 
$\omega^{(0)}_{p}$ the function $\sin(\omega^{(0)}_{p}n)$ can be reduced to 
a combination of basic harmonics of $\Omega_{i}n$, one concludes that in the 
generic case in which every $l_{i}/L_{0}$ is an irrational number, the phases 
\begin{equation}
x_{i,n}=\left[\Omega_{i}n\right]_{\mathop{\rm mod}2\pi}.
\label{xi}
\end{equation}
in each term in (\ref{omegas}) will generate random outputs, uniformly distributed in 
the interval $[0,2\pi]$. Hence, in the context of studying the statistical properties 
of the eigenvalue sequence, the argument of every factor $\sin(\omega^{(0)}_{p}n)$ in 
(\ref{kn}) can be treated as a function
\begin{equation}
\sin \left(\omega^{(0)}_{p}n\right) \rightarrow \sin \left(
m^{(0)}_{p,1}x_{1}+...+m^{(0)}_{p,N_{B}}x_{N_{B}}\right) =\sin \left(\vec m^{(0)}_{p}\vec{x}\right) ,
\end{equation}
of $N_{B}-1$ independent random variables $x_{i}$, that are distributed in the 
interval $[0,2\pi ]$. Hence the set of the deviations of the eigenvalues from the 
average is statistically described by a series or random inputs corresponding 
to the periodic orbit expansion (\ref{deltakn})
\begin{equation}
\delta_{x}^{(0)}=-\frac{2}{\pi}\sum_{p}\frac{A_{p}^{(0)}}{\omega^{(0)}_{p}}\sin
\left(\frac{\omega^{(0)}_{p}}{2}\right) \sin \left(\vec m^{(0)}_{p}\vec{x}\right).
\label{deltax}
\end{equation}
The maximal amplitude of an input corresponding to a periodic orbit $p$ coincides 
with the amplitude of the same orbit's contribution into the exact periodic orbit 
sum (\ref{kn}).

It is now a straightforward task to obtain the distribution of $\delta_{n}^{(0)}$ 
via 
\begin{equation}
P_{\delta}^{(0)}\left(\delta^{(0)}\right) =\int_{0}^{2\pi}...\int_{0}^{2\pi}
\delta \left(\delta^{(0)}+\sum_{p}C_{p}^{(0)}\sin\left(\vec m^{(0)}_{p}\vec{x}\right)
\right)\prod_{i=1}^{N_{B}-1} \frac{dx_{i}}{2\pi}.
\end{equation}
where
\begin{equation}
C_{p}^{(0)}=
\frac{2}{\pi}\frac{A_{p}^{(0)}}{\omega^{(0)}_{p}}\sin\left(\frac{\omega^{(0)}_{p}}{2}\right).
\label{cp0}
\end{equation}
Using the exponential representation of the $\delta $-functional, one has 
\begin{eqnarray}
P_{\delta}^{(0)}\left(\delta^{(0)}\right) =\int dke^{ik\delta^{(0)}}
F_{\delta}^{(0)}\left(k,C_{p}^{(0)}\right) , 
\label{distribr}
\end{eqnarray}
where the characteristic function $F_{\delta}^{(0)}\left(k,C_{p}^{(0)}\right)$ is 
defined explicitly via the periodic orbits and graph parameters, 
\begin{equation}
F_{\delta}^{(0)}\left(k,C_{p}^{(0)}\right) =\int_{0}^{2\pi}...\int_{0}^{2\pi}
e^{ik\sum_{p}C_{p}^{(0)}\sin \left(\vec m^{(0)}_{p}\vec{x}\right)}\prod_{i=1}^{N_{B}-1}
\frac{dx_{i}}{2\pi}.
\label{distreg}
\end{equation}
As in the case of the series expansion for $k_{n}$, a finite order ($m$th) correction 
to the exact result is obtained by considering the orbits that involve the same number 
$|m^{(0)}_{p}|=m^{(0)}_{p,1}+...+m^{(0)}_{p,N_{B}}$ of vertex scatterings.

In general, the statistical properties of a spectral quantity $z_{n}$ that has a periodic 
orbit series expansion
\begin{equation}
z_{n}^{(0)}=
f^{(0)}_{z}-\sum_{p}c_{p}^{(0)}\cos\left(\omega^{(0)}_{p}n+\varphi^{(0)}_{p}\right),
\label{zn0}
\end{equation}
where $\varphi^{(0)}_{p}$ is an $n$-independent phase and $f^{(0)}_{z}$ is a shift term 
(see below), are described by the random sequence 
\begin{equation}
z_{x}^{(0)}=
f^{(0)}_{z}-\sum_{p}c_{p}^{(0)}\cos\left(\vec m^{(0)}_{p}\vec{x}+\varphi^{(0)}_{p}\right).
\label{zx0}
\end{equation}
The corresponding characteristic function of probability distribution for $z$ will have 
the form
\begin{equation}
F_{z}^{(0)}(k) =\int_{0}^{2\pi}...\int_{0}^{2\pi}
e^{ik\sum_{p}c_{p}^{(0)}\cos \left(\vec m^{(0)}_{p}\vec{x}+\varphi^{(0)}_{p}\right)}dx,
\label{Fz}
\end{equation}
where $dx=\prod_{i=1}^{N_{B}-1}\frac{dx_{i}}{2\pi}$.

Another case of spectral characteristics that can be treated by this approach 
is provided e.g. by the separation between two eigenvalues, $k_{n+m}$ and $k_{n}$, 
for a fixed $m$. This quantity is defined by the expansion
\begin{equation}
s_{n,m}^{(0)}=\frac{\pi}{L_{0}}m-\sum_{p}D_{p,m}^{(0)}\cos\left(\omega^{(0)}_{p}
n+\omega^{(0)}_{p}\frac{m}{2}\right),
\label{snm0}
\end{equation}
with
\begin{equation}
D_{p,m}^{(0)}=\frac{4}{L_{0}}\frac{A_{p}^{(0)}}{\omega^{(0)}_{p}}\sin \left(
\frac{\omega^{(0)}_{p}}{2}\right) \sin \left(\frac{\omega^{(0)}_{p}m}{2}\right),
\label{dpm}
\end{equation}
which gives rise to the random series
\begin{equation}
s_{m,x}^{(0)}=\frac{\pi m}{L_{0}}-\sum_{p}D_{p}^{(0)}\cos \left(\vec m^{(0)}_{p}
\vec{x}-\frac{m\omega^{(0)}_{p}}{2}\right).
\label{smx0}
\end{equation}
The distribution of $s_{m}^{(0)}$ can be obtained as 
\begin{equation}
P_{s_{m}}^{(0)}\left(s\right) =\int dke^{ik\left(s-\frac{\pi m}{L_{0}}
\right)}F_{s_{m}}^{(0)}\left(k,D_{p,m}^{(0)}\right)  
\label{psm0}
\end{equation}
where $F_{s_{m}}^{(0)}\left(k,D_{p,m}^{(0)}\right)$ is defined according
to (\ref{Fz}). In the case $m=1$, these formulae describe the nearest
neighbors distribution. Clearly, for the difference between the fluctuations
themselves, $\sigma_{n,m}^{(0)}=\delta_{n+m}-\delta_{n}$, the distribution is 
$P_{\sigma_{m}}^{(0)}(\sigma) =P_{s_{m}}^{(0)}\left(s+\frac{\pi m}{L_{0}}\right)$.

One could also study the mean of the two neighboring deviations, 
$\xi_{n}^{(0)}=\left(\delta_{n}^{(0)}+\delta_{n-1}^{(0)}\right)/2$.
This characteristics can be used to generate the separators directly 
from $k_{n}^{(0)}$, $k_{n}^{(1)}=n+\xi_{n}^{(0)}$ independently of the 
properties of the spectral determinant. The corresponding series has the
expansion coefficients
\begin{equation}
E_{p}^{(0)}=
\frac{1}{\pi}\frac{A_{p}^{(0)}}{\omega^{(0)}_{p}}\sin \left(\omega^{(0)}_{p}\right)
\label{ep0}
\end{equation}
and phases $\varphi^{(0)}_{p}=\frac{\pi}{2}-\frac{\omega^{(0)}_{p}}{2}$. 
The energy fluctuations, $\delta E_{n}$, have the expansion coefficients
\begin{equation}
H_{p}^{(0)}=\frac{2\pi}{L_{0}^{2}}\frac{A_{p}^{(0)}}{\omega^{(0)}_{p}}
\left(\frac{2}{\omega^{(0)}_{p}}\sin \left(\frac{\omega^{(0)}_{p}}{2}\right) 
-\cos \frac{\omega^{(0)}_{p}}{2}\right),
\label{hp0}
\end{equation}
phases $\varphi^{(0)}_{p}=0$ and $f_E=\frac{\pi^{2}}{12L_{0}^{2}}$, etc.

These quantities can be used to find the periodic orbit expansions for higher order 
statistics, such as the correlator $\left\langle\delta_{n}^{(0)}\delta_{n+m}^{(0)}
\right\rangle$ or the autocorrelation function,
\begin{equation}
R_{2}(x)=\frac{\pi^{2}}{L_{0}^{2}}\left\langle \rho \left(k+\frac{x}{2}\right)
\rho \left(k-\frac{x}{2}\right) \right\rangle=
\frac{\pi}{L_{0}}\lim_{N\rightarrow \infty}\frac{1}{N}
\sum_{n=1}^{N}\sum_{m\neq 0}\delta \left(k_{n+m}-k_{n}+x\right).
\label{autocor}
\end{equation}
which defines the probability to find a new level at a distance $x\neq 0$ from a 
given old one, whether or not these levels are nearest neighbors. The Fourier image 
of $R_{2}(x)$, the form factor, $K_{2}(\tau)$ is given by 
\begin{eqnarray}
K_{2}(\tau )=\frac{\pi}{L_{0}}\lim_{N\rightarrow \infty}\frac{1}{N}
\sum_{n=1}^{N}\sum_{m=1}^{N}e^{i\left(k_{n+m}-k_{n}\right) \tau}=
\frac{\pi}{L_{0}}\left\langle\sum_{m}e^{-i\left(k_{n+m}-k_{n}\right)\tau}
\right\rangle_{n}.
\label{k2k}
\end{eqnarray}
For the regular graphs with $k_{n}$ given by (\ref{kn}), the averaging over $n$ can be 
performed via 
\begin{eqnarray}
K_{2}(\tau )=\frac{\pi}{L_{0}}\lim_{N\rightarrow \infty}\frac{1}{N}
\sum_{n=1}^{N}\sum_{m}e^{i\left(-\frac{\pi}{L_{0}}m+\sum_{p}D_{p,m}^{(0)}
\cos \left(\omega^{(0)}_{p}\left(n+\frac{m}{2}\right) \right) \right) \tau}\cr=
\frac{\pi}{L_{0}}\sum_{m}e^{-i\frac{\pi m}{L_{0}}\tau}\int
e^{i\sum_{p}\tau D_{p,m}^{(0)}\cos \left(\vec m^{(0)}_{p}\vec{x}+\omega^{(0)}_{p}
\frac{m}{2}\right)}\prod_{i}\frac{dx_{i}}{2\pi}.
\end{eqnarray}
The latter integral is defined by (\ref{psm0}), so 
\begin{equation}
K_{2}(\tau )=\frac{\pi}{L_{0}}\sum_{m=1}^{\infty}e^{-i\frac{\pi m}{L_{0}}
\tau}F_{s_{m}}^{(0)}\left(\tau ,D_{p,m}^{(0)}\right).
\label{k2tau}
\end{equation}
Formula (\ref{k2tau}) can also be obtained directly from (\ref{k2k}) by 
averaging over the random variable $s_{m}^{(0)}$ using the distribution (\ref{psm0}),
\begin{eqnarray*}
K_{2}(\tau ) &=&\frac{\pi}{L_{0}}\left\langle\sum_{m}e^{-i\left(k_{n+m}-k_{n}
\right)\tau}\right\rangle_{n}=\frac{\pi}{L_{0}}\sum_{m}
\left\langle e^{-is_{m}^{(0)}\tau}\right\rangle_{s_{m}}.
\end{eqnarray*}
Hence $R_{2}(x)$ is given by
\begin{eqnarray}
R_{2}(x)=\frac{\pi}{L_{0}}\sum_{m}\int e^{i\tau \left(x-\frac{\pi m}{L_{0}}\right)}
F_{s_{m}}^{(0)}\left(\tau ,D_{p,m}^{(0)}\right) d\tau =\frac{\pi}{L_{0}}\sum_{m}
P_{s_{m}}^{(0)}(x),
\end{eqnarray}
as the sum of the probabilities that the two eigenvalues separated by the interval 
$x$ have $m-1$ other eigenvalues in-between.

It should be emphasized that all the probability distributions above are obtained 
in the context of the standard periodic orbit theory framework. All the distributions 
for the regular level fluctuations derived in this section are closed, self contained 
expressions, defined in terms of the periodic orbits and graph parameters. It is also 
important to notice that the statistical properties of some (especially some regular) 
quantum graphs, despite being strongly stochastic in the classical regime, deviate from 
the universal Wignerian distributions predicted by the RMT. 
However these cases, as well as the irregular graphs discussed below, are equally well 
described via statistical description of the periodic orbit expansion series for the 
spectral sequences.

\section{Spectral expansions for irregular graphs}

As mentioned in the Section II, one can find the roots of $\Delta^{(j-1)}(k)$ by using 
the density $\rho^{(j)}(k)$ and the separators $\hat{k}_{n}^{(j)}$ in formula (\ref{seps}), 
which yields 
\begin{eqnarray}
\hat{k}_{n}^{(j-1)}=
\hat{k}_{n}^{(j)}N^{(j-1)}\left(\hat{k}_{n}^{(j)}\right) -
\hat{k}_{n-1}^{(j)}N^{(j-1)}\left(\hat{k}_{n-1}^{(j)}\right) -
\int_{\hat{k}_{n-1}^{(j)}}^{\hat{k}_{n}^{(j)}}N^{(j-1)}(k)dk.
\end{eqnarray}
Every staircase function $N^{(j)}(k)$ can be decomposed into the average and the oscillating 
parts, $N^{(j)}(k)=\bar{N}^{(j)}(k)+\delta N^{(j)}(k)$, where the average integrated density 
for every $j$ is
\begin{equation}
\bar{N}^{(j)}(k)=\frac{L_{0}}{\pi}k-\frac{1}{2}=\bar{N}(k).
\label{Njave}
\end{equation}
The bootstrapping (\ref{bootstrap}) of $\hat{k}_{n}^{(j-1)}$ by $\hat{k}_{n}^{(j)}$ 
(or $N^{(j-1)}(k)$ by $N^{(j)}(k)$, see Fig.\ref{Bootstrapping}) implies that 
\begin{equation}
N^{(j-1)}\left(\hat{k}_{n}^{(j)}\right) =n.
\label{jpierce}
\end{equation}
Using (\ref{Njave}), (\ref{jpierce}) and writing $\hat{k}_{n}^{(j)}$ in the form 
\begin{equation}
\hat{k}_{n}^{(j)}=\frac{\pi}{L_{0}}\left(n+\delta_{n}^{(j)}\right) ,
\label{knjdecomposition}
\end{equation}
we get for the $j-1$ generation of the separators 
\begin{eqnarray}
\hat{k}_{n}^{(j-1)}=\frac{\pi n}{L_{0}}+\frac{1}{2}\frac{\pi}{L_{0}}\left(\delta_{n}^{(j)}
-\delta_{n-1}^{(j)}\right) -\frac{1}{2}\frac{\pi}{L_{0}}\left((\delta_{n}^{(j)})^{2}
-(\delta_{n-1}^{(j)})^{2}\right) -\int_{\hat{k}_{n-1}^{(j)}}^{\hat{k}_{n}^{(j)}}\delta 
N^{(j-1)}(k)~dk.
\end{eqnarray}
The harmonic series expansion for $\delta N^{(j)}(k)$, 
\begin{equation}
\delta N^{(j)}(k)=\frac{1}{\pi}\mathop{\rm Im}\sum_{p}A_{p}^{(j)}e^{iL^{(j)}_{p}k},
\label{series}
\end{equation}
allows to compute the integral in (\ref{jfluct}) explicitly, which yields 
the fluctuating part of $\hat{k}_{n}^{(j-1)}$ via an expansion similar to 
(\ref{deltakn}), 
\begin{eqnarray}
\delta_{n}^{(j-1)}=f^{(j-1)}_{\delta}\left(\delta_{n}^{(j)},\delta_{n-1}^{(j)}\right) 
-\sum_{p}C_{p}^{(j-1)}\left(\delta_{n}^{(j)},\delta_{n-1}^{(j)}\right)
\sin\left(\omega^{(j-1)}_{p}n+\varphi^{(j-1)}_{p}\left(\delta_{n}^{(j)},\delta_{n-1}^{(j)}\right)
\right),
\label{jfluct}
\end{eqnarray}
where now the expansion coefficients, 
\begin{equation}
C_{p}^{(j-1)} =
\frac{2}{\pi}\frac{A_{p}^{(j-1)}}{\omega^{(j-1)}_{p}}\sin \frac{\omega^{(j-1)}_{p}}{2}
\allowbreak \left(\delta_{n}^{(j)}-\delta_{n-1}^{(j)}+1\right),
\label{cpj}
\end{equation}
the ``zero shift'' term,
\begin{equation}
f^{(j-1)}_{\delta}=\frac{1}{2}
\left(\delta_{n}^{(j)}-\delta_{n-1}^{(j)}\right) -\frac{1}{2}\left(
(\delta_{n}^{(j)})^{2}-(\delta_{n-1}^{(j)})^{2}\right),
\label{fdelta}
\end{equation}
and the phases,
\begin{equation}
\varphi^{(j-1)}_{p}=
\frac{\delta_{n}^{(j)}+\delta_{n-1}^{(j)}-1}{2}\omega^{(j-1)}_{p},
\label{fipj}
\end{equation}
are functions of the fluctuations $\delta_{n}^{(j)}$ and $\delta_{n-1}^{(j)}$ 
on the previous level of the hierarchy. In the particular case when $r=0$, 
$\delta_{n}^{(1)}=\delta_{n-1}^{(1)}=1/2$, (\ref{jfluct}) coincides with the 
oscillating part of (\ref{kn}).

The equation (\ref{jfluct}) shows that the fluctuations $\delta_{n}^{(j-1)}$ 
have two sources. In addition to the oscillations induced by the periodic orbit 
sum in (\ref{jfluct}), there are also oscillating contributions produced by 
$\delta_{n}^{(j)}$ and $\delta_{n-1}^{(j)}$ that bring in the oscillations from 
all the previous levels of the hierarchy. This equation will later be used to 
produce the exact statistical distribution for the fluctuations $\delta_{n}^{(j)}$ 
at each level of the hierarchy.

As an example of a case where the coefficients $A^{(j)}_{p}$ can be obtained directly, 
one can use a simple example of a 2-bond regular graph, discussed in \cite{Prima,Opus}. 
Although the 2-bond graph is a strictly regular system, which does not require auxiliary 
separating sequences for obtaining its spectrum, it is nevertheless  useful to use this 
case to have an immediate illustration of the explicit form of the coefficients $A^{(j)}_{p}$. 

As shown in \cite{Nova,Prima,Opus}, the spectral equation in this case has the form 
\begin{equation}
\sin((l_{1}+l_{2})k)-r\sin((l_{1}-l_{2})k)=0,
\label{2bond}
\end{equation}
where $r$ is the reflection coefficient at the middle vertex and $l_{1}$ and $l_{2}$ are 
the two bond lengths. This equation was obtained in \cite{Nova,Prima,Opus} using scattering
quantization method \cite{QGT1,QGT2} as $\Delta(k)=\det(1-S(k))=0$, where scattering matrix
$S(k)\equiv S^{(0)}(k)$ in this case is
\begin{eqnarray}
S^{(0)}(k)=\pmatrix{ 
0 & -e^{il_{1}k} & 0 & 0 \cr
re^{il_{1}k} & 0 & 0 & te^{il_{2}k} \cr
te^{il_{1}k} & 0 & 0 & -re^{il_{2}k}\cr
0 & 0 & -e^{il_{2}k} & 0}.
\label{coeff0}
\end{eqnarray}
The unitarity of $S(k)$ is guaranteed by the ``flux conservation'' relationship between 
the reflection and the transmission coefficient, $t^{2}+r^{2}=1$. The expansion of the 
spectral determinant \cite{QGT1,QGT2,Opus}, yields the exact periodic orbit expansion 
for $N(k)$,
\begin{eqnarray}
N(k)=\bar N(k)- \frac{1}{ \pi}{\rm Im} \sum_{p,\nu} \frac{1}{\nu}
\left[(-1)^{\chi(p)}t^{2\tau(p)}r^{\sigma(p)}\right]^{ \nu }
\, e^{i\nu L_{p}k}
\label{nexpansion}
\end{eqnarray}
where $\nu$ is the multiple traversal index for the prime periodic trajectory $p$, $\sigma(p)$ 
and $2\tau(p)$ are the numbers of reflections and transmissions for $p$ at the middle vertex, 
and the factor $(-1)^{\chi(p)}$ defines the Maslov index \cite{Nova}. The action length of $p$ 
is $L_{p}=m_{p,1}l_{1}+m_{p,2}l_{2}$, according to the number of times, $m_{p,1}$  and $m_{p,2}$, 
it traverses the bonds $l_{1}$ and $l_{2}$. One can notice that the equation for the separating 
points $\hat k^{(1)}_{n}$,
\begin{eqnarray}
\cos((l_{1}+l_{2})k)-r\omega\cos((l_{1}-l_{2})k)=0,
\label{derivative}
\end{eqnarray}
where $\omega=(l_{1}-l_{2})/(l_{1}+l_{2})<1$, reduces to the original equation (\ref{2bond}),
if $r\rightarrow r^{(1)}=\omega r$ and $l_{1}k\rightarrow l_{1}k+\frac{\pi}{2}$. Therefore, 
the expansion for $N^{(1)}(k)$ can be obtained from $\det(1-S^{(1)}(k))$, where unitary matrix 
$S^{(1)}(k)$ is obtained via a two parameter deformation of the unitary matrix $S^{(0)}(k)$,
\begin{eqnarray}
S^{(1)}=\pmatrix{ 
0 & -e^{il_{1}k+\frac{\pi}{2}} & 0 & 0 \cr
r^{(1)}e^{il_{1}k+\frac{\pi}{2}} & 0 & 0 & t^{(1)}e^{il_{2}k} \cr
t^{(1)}e^{il_{1}+\frac{\pi}{2}} & 0 & 0 & -r^{(1)}e^{il_{2}k}\cr
0 & 0 & -e^{il_{2}k} & 0}.
\label{coeff1}
\end{eqnarray}
with $r^{(1)}=\omega r$ and $t^{(1)}=\sqrt{1-\left(r^{(1)}\right)^2}$. The weight coefficients 
$A^{(1)}_{p}$ are now explicitly defined via the products of the matrix elements of $S^{(1)}(k)$, 
just as the $A^{(0)}_{p}$ coefficients were defined via $S^{(0)}(k)$ in (\ref{nexpansion}).

Clearly, all even order derivatives of the spectral equation (\ref{2bond}) for $j=2w$ 
have the form $\sin((l_{1}+l_{2})k)-r\omega^{2w}\sin((l_{1}-l_{2})k)=0$, so with the 
replacement $r^{(2w)}=\omega^{(2w)}r$, $t^{(2w)}=\sqrt{1-\left(r^{(2w)}\right)^2}$, the 
form of the coefficients $A^{(2w)}_{p}$ is structurally same as the one found in 
(\ref{nexpansion}). The odd degree derivatives $A^{(2w+1)}_{p}$ are produced by the 
expansion of the determinant $\det(1-S^{(2w+1)}(k))$ analogous to (\ref{coeff1}). 
It should be mentioned however, that in general the task of obtaining the exact form 
of the coefficients $A^{(j)}_{p}$ and frequencies $\omega^{(j)}_{p}$ for $j\neq 0$ is 
not as straightforward as in this simple case and requires a more detailed analysis. In 
particular, the harmonic expansions such as (\ref{jfluct}) may include bond combinations
that do not correspond to connected periodic orbits.

Using the expansion for the fluctuations $\delta^{(j-1)}_{n}$, one can find the harmonic
expansions for other spectral characteristics. For example, the $m$-neighbor difference, 
$s_{n,m}^{(j-1)}=m+\delta_{n+m}^{(j-1)}-\delta_{n}^{(j-1)}$, the expansion is    
\begin{eqnarray}
s_{n,m}^{(j-1)}=
f^{(j-1)}_{s}\left(s_{n-1,m}^{(j)},s_{n,m}^{(j)},s_{n,m-1}^{(j)},\xi_{n}^{(j)}\right)
+\sum_{p}D_{p,m}^{(j-1)}\cos\left(n\omega^{(j-1)}_{p}+\varphi^{(j-1)}_{p}\right), 
\label{smnj}
\end{eqnarray}
where
\begin{eqnarray}
f^{(j-1)}_{s}=s_{n,m}^{(j)}+\frac{1}{2}\left(s_{n,m}^{(j)}-s_{n,m-1}^{(j)}\right)
\left(2m-s_{n,m}^{(j)}-s_{n-1,m}^{(j)}\right) +\xi_{n}^{(j)}\left(s_{n-1,m}^{(j)}
-s_{n,m}^{(j)}\right)
\label{fs}
\end{eqnarray}
and $\xi_{n}^{(j)}=\frac{1}{2}\left(\delta_{n}^{(j)}+\delta_{n-1}^{(j)}\right)$. 
The harmonic expansion coefficients in this case are
\begin{eqnarray}
D_{p,m}^{(j-1)}=\frac{A_{p}^{(j-1)}}{\omega^{(j-1)}_{p}}
\Big(\sin^{2}\frac{\omega^{(j-1)}_{p}}{2}s_{n,m}^{(j)}+\sin ^{2}\frac{\omega^{(j-1)}_{p}}{2}
s_{n-1,m}^{(j)}
\cr
-2\left(\sin \frac{\omega^{(j-1)}_{p}}{2}s_{n,m}^{(j)}\sin \frac{\omega^{(j-1)}_{p}}{2}
s_{n-1,m}^{(j)}\right) \cos \frac{\omega^{(j-1)}_{p}}{2}
\left(2s_{n,m-1}^{(j)}-s_{n,m}^{(j)}-s_{n-1,m}^{(j)}\right) \Big)^{1/2}
\label{dpmj}
\end{eqnarray}
In case $m=1$ the system of equations (\ref{smnj}) yields the nearest neighbor 
distances between of the $j-1$ level separators.
%%%%%%%%%%%%%%%%%%%%%%%%%%%%%%%%%%%%%%%%%%%%%%%%%%%%%%%%%%%%%%%%%%%%%%
\begin{figure}[tbp]
\begin{center}
\includegraphics{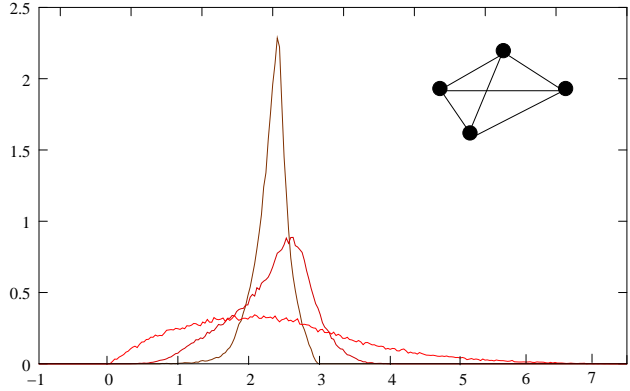}
\end{center}
\caption{Histogram of the nearest neighbor separations $s^{(0)}_n$ (bottom 
distribution), $s^{(1)}_n$ and $s^{(2)}_n$ (top distribution) for the fully 
connected quadrangle graph (top right corner) of the irregularity degree 3, 
obtained for 80,000 roots of the corresponding spectral equations. The maximal 
nearest neighbor separation for $j=0$ in this case is $s_{max}=8.68$, while 
regular cell size is $\pi/S_{0}=2.28$.}
\label{Separations3}
\end{figure}
%%%%%%%%%%%%%%%%%%%%%%%%%%%%%%%%%%%%%%%%%%%%%%%%%%%%%%%%%%%%%%%%%%%%%%

The nearest neighbor average, $\xi^{(j)}_{n}$, that appears e.g. in (\ref{dpmj})
has the expansion
\begin{eqnarray}
\xi_{n}^{(j-1)}=f^{(j-1)}_{\xi}\left(s_{n}^{(j)},s_{n-1}^{(j)},\xi_{n}^{(j)}\right) -
\sum_{p}E_{p}^{(j-1)}\left(s_{n}^{(j)},s_{n-1}^{(j)}\right) 
\sin \left(\omega^{(j-1)}_{p}n+\varphi^{(j-1)}_{p}\right),
\label{ksij}
\end{eqnarray}
where 
\begin{eqnarray*}
f^{(j-1)}_{\xi} &=&-\frac{1}{4}
\frac{L_{0}^{2}}{\pi^{2}}\left(s_{n}^{(j)}+s_{n-1}^{(j)}-\frac{2\pi}{L_{0}}
\right) \left(s_{n}^{(j)}-\frac{2\pi}{L_{0}}+\frac{2\pi}{L_{0}}\xi_{n-1}^{(j)}
\right) 
\cr 
E_{p}^{(j-1)}&=&\frac{1}{L_{0}}
\frac{A_{p}^{(j-1)}}{\omega^{(j-1)}_{p}}\sin \frac{L^{(j-1)}_{p}}{2}
\left(s_{n}^{(j)}+s_{n-1}^{(j)}\right) 
\cr 
\varphi^{(j-1)}_{\xi, p}&=&
\left(\frac{L_{0}}{2}s_{n}^{(j)}+\xi_{n-1}^{(j)}-\frac{3}{2}\right)\omega^{(j-1)}_{p}.
\end{eqnarray*}
The expansions of the kind (\ref{jfluct}), (\ref{smnj}) and (\ref{ksij}), provide
an exact description of the propagation of the spectral characteristics across the 
hierarchy in terms of the geometry of the graph. For the graphs of high degree of 
irregularity, the equations (\ref{jfluct}), (\ref{smnj}), (\ref{ksij}), etc., can 
be considered as discretizations of nonlinear differential equations for continuous 
functions $\delta_{n}^{(j)}\sim \delta(n,j)$, $s_{n,m}^{(j)}\sim s_{m}(n,j)$, 
$\xi_{n}^{(j)}\sim \xi(n,j)$, etc.

\section{Fluctuation Statistics}

As demonstrated in the previous section, the fluctuations $z_{n}^{(j)}$
(e.g. (\ref{jfluct}), (\ref{smnj}), (\ref{ksij})) at the $j$th level of 
the hierarchy depend on the fluctuations on all the previous levels as well 
as on the oscillations introduced by the harmonic terms at the level $j$, 
\begin{equation}
z_{n}^{(j-1)}=f^{(j-1)}\left(z_{n}^{(j)}\right) -\sum_{p}c_{p}^{(j-1)}
\left(z_{n}^{(j)}\right) \cos \left(\omega^{(j-1)}_{p}n+\varphi^{(j-1)}_{p}
\left(z_{n}^{(j)}\right) \right),
\label{accumn}
\end{equation}
As in the regular case considered in Chapter III, probabilistic description of the 
sequence $z^{(j)}$ at each $j$ is obtained by considering $z^{(j)}$ as a function of 
a random vector $\vec{x}$, generated by the sequence 
$x_{i}=\left[\Omega_{i}n\right]_{\mathop{\rm mod}2\pi}$,
\begin{equation}
z_{x}^{(j-1)}=f^{(j-1)}\left(z_{x}^{(j)}\right)-\sum_{p}c_{p}^{(j-1)}\left(z_{x}^{(j)}\right) 
\cos \left(\vec m^{(j-1)}_{p}\vec{x}+\varphi^{(j-1)}_{p}\left(z_{x}^{(j)}\right)\right).
\label{zxj}
\end{equation}
Since the index $n$ and the ``bond frequencies'' $\Omega_{i}$ are the same 
for all the $z_{n}^{(j)}$ expansions across the hierarchy, the random variable 
$\vec x$ is the same in all the harmonic expansion terms in the stochastic 
series (\ref{zxj}), $j=0,\ ...,\ r$. This implies however, that unlike the 
regular case, the harmonic oscillations at the higher levels of the hierarchy 
($j<r$) are not the only source of randomness in the series (\ref{zxj}). 
Additional ``noise'' at the level $j$ is injected into the equation (\ref{zxj}) 
by the variables $z_{x}^{(j)}$, that introduce the fluctuations from all the 
previous levels of the hierarchy into its $j-1$ level. The expansions (\ref{zxj}) 
allow to obtain the probability distributions for each of the $z^{(j)}$s from
\begin{equation}
P_{z}^{(j-1)}(z^{(j-1)})=\int \delta \left(z^{(j-1)}-z^{(j-1)}\left({z_{x}^{(j)}}
\left({z_{x}^{(j+1)}}(...),x\right),x\right) \right) dx,
\label{probability}
\end{equation}
where $dx=\prod\limits_{i}\frac{dx_{i}}{2\pi}$. The ``nested'' structure 
of the $z_{x}^{(j-1)}$ in (\ref{probability}) can be unfolded in the form 
\begin{eqnarray}
P_{z}^{(j-1)}(z^{(j-1)}) &=&\int \delta \left(z^{(j-1)}-z^{(j-1)}
\left({z_{x}^{(j)},x}\right) \right)\cr&&\times\delta \left(z_{i}^{(j)}-z_{i}^{(j)}
\left({z_{x}^{(j+1)},x}\right)\right)\cr &&...\times \delta 
\left(z_{i}^{(r)}-z_{i}^{(r)}(x)\right)\prod_{j} \prod_{i} z^{(j)}_{i}dx.
\label{pjdeltas}
\end{eqnarray}
in which the chain of the delta functionals above can be understood as the 
conditional probability densities for obtaining the value $z^{(j-1)}$, given 
the values of $\vec x$ and $z^{(j)}$, $z^{(j+1)}$, ... , $z^{(r)}$. Here the 
index $i$ runs over the number of the elements $z^{(j)}_{i}$ that appear in 
the expansion (\ref{accumn}) of the corresponding quantity $z^{(j-1)}$. For 
example, the expansion of $\delta^{(j-1)}_{n}$ depends on $\delta^{(j)}_{n-1}$ 
and $\delta^{(j)}_{n}$, so in this case index $i$ takes two values, $i=1,2$. 
In the following this index will be omitted for conciseness of notations.

The hierarchical organization of the fluctuations suggests a natural way of 
approximate evaluation of the expression (\ref{pjdeltas}). Since for every 
$j$ the function $z^{(j)}_{x}$ is a complex, rapidly oscillating function of 
$x$, and since every $z^{(j-1)}$ characteristics depends explicitly only on 
the previous level sequence $z^{(j)}$, it is natural to consider a simple 
approximation to (\ref{pjdeltas}), in which $z^{(j)}$ in the equations (\ref{zxj})
are treated as independent random variables distributed according to $P_{z}^{(j)}$,
\begin{eqnarray}
P_{z}^{(j-1)}(z^{(j-1)})=\int \delta \left(z^{(j-1)}-z^{(j-1)}
\left({z^{(j)},x}\right)\right)P^{(j)}_{z}dz^{(j)}dx.
\label{approx}
\end{eqnarray}
A more formal and elaborate argument that will not be discussed here in detail 
is based on inverting the dependence $z^{(j-1)}=z^{(j-1)}(x)$ for each $j$ in 
the arguments of the delta functionals in (\ref{pjdeltas}) (i.e. in effect using 
Bayes relationships $P(z|x)P(x)=P(x|z)P(z)$). It is clear that due to the complex 
oscillatory behavior of the expansions $z^{(j-1)}(x)$, one can approximate each of 
the $P_{z}^{(j)}(x|z)$ distributions by a uniform distribution, which effectively 
corresponds to introducing separate $x$-variables for each level $j$, and leaves, 
after intermediate integrations, the distribution $P^{(j)}_{z}$ in (\ref{approx}).

Such approach takes a natural advantage the hierarchical organization of the
fluctuations produced by the sequences $z^{(j)}$. Since the distribution $P_{z}^{(r)}$ 
at the regular level can be computed directly, the distribution $P_{z}^{(r-1)}$, 
$P_{z}^{(r-2)}$, ..., can be computed in sequence, from the previous levels of the 
hierarchy to the next. The final distributions $P_{z}^{(0)}$ will apply to the physical 
characteristics of the spectrum.

As an example, one can consider the transition from the level $j$ to the level $j-1$ 
of the distributions for the deviations from the average. The $j$th level fluctuations 
$\delta_{n}^{(j)}$ and $\delta_{n-1}^{(j)}$ in (\ref{jfluct}), can now be considered 
as random variables, $\delta_{1}$ and $\delta_{2}$, distributed according to 
$P_{\delta}^{(j)}\left(\delta\right)$. Hence for the density $P_{\delta}^{(j-1)}(\delta)$ 
one can write 
\begin{eqnarray}
P_{\delta}^{(j-1)}\left(\delta \right) =\int d\delta_{1}d\delta_{2}dx 
P_{\delta}^{(j)}\left(\delta_{1}\right) P_{\delta}^{(j)}\left(\delta_{2}\right)
\delta{\Bigg(}\delta -f^{(j-1)}_{\delta}\left(\delta_{1}^{(j)},\delta_{2}^{(j)}\right) 
\cr
+\sum_{p}C_{p}^{(j-1)}\left(\delta_{1}^{(j)},\delta_{2}^{(j)}\right) \sin 
\left(\vec m^{(j-1)}_{p}\vec{x} +\varphi^{(j-1)}_{p}\left(\delta_{1}^{(j)},
\delta_{2}^{(j)}\right) \right){\Bigg)}.
\end{eqnarray}
Exponentiation of the delta function produces 
\begin{equation}
P^{(j-1)}(\delta) =\int dke^{ik\delta}\left\langle F_{\delta}^{(j-1)}
\left(k,C_{p}^{(j-1)}\right)\right\rangle_{\Omega^{(j)}},
\label{avedeltaj}
\end{equation}
where the function
\begin{equation}
F_{\delta}^{(j-1)}(k) =\int_{0}^{2\pi}...\int_{0}^{2\pi}e^{ik
\sum_{p}C_{p}^{(j-1)}\left(\delta_{1},\delta_{2}\right) 
\sin\left(\vec m^{(j-1)}_{p}\vec{x}+\varphi^{(j-1)}_{p}
\right)}dx,
\label{Fdeltaj}
\end{equation}
produced by the integral over $x_{i}$s is reminiscent of the (\ref{distreg}).
The parenthesis $\left\langle \ast ,\ast \right\rangle_{\Omega^{(j)}}$
denote averaging of the $F_{\delta}^{(j)}(k)$ with the weight 
\begin{equation}
\Omega^{(j)}(\delta^{(j)}_{1},\delta^{(j)}_{2},k) =
e^{-ikf^{(j-1)}_{\delta}\left(\delta_{1}^{(j)},\delta_{2}^{(j)}\right)}P_{\delta}^{(j)}
\left(\delta_{1}\right)P_{\delta}^{(j)}\left(\delta_{2}\right),
\end{equation}
which corresponds to averaging over the ``separator disorder'', which yields the 
characteristic function of the distribution (\ref{avedeltaj}). In case if 
$P_{\delta}^{(r+1)}=\delta \left(\delta^{(r+1)}-\frac{1}{2}\right)$ (ordered
separators at $j=r$), one recovers the regular level distribution (\ref{distribr}).
For $j<r$, obtaining the probability distributions $P^{(j-1)}_{\delta}(\delta)$
for requires an additional averaging over the disorder produced by the fluctuating 
sequences of the separators $\hat k^{(j)}_{n}$.

The expression for other spectral characteristics have similar structure. 
In the case of the $m$-th neighbor separation statistics, the probability 
distribution for $s_{m}^{(j-1)}$ is given by 
\begin{eqnarray}
P_{m}^{(j-1)}(s)=
\int e^{ik\left(s-m\right)}\left\langle F_{s_{m}}^{(j)}\right
\rangle_{\Phi^{(j)}}dk, 
\label{avesnmj}
\end{eqnarray}
where the ``characteristic exponential'' $F_{s_{m}}^{(j)}$ generated by 
the harmonic series $\delta s_{x,m}^{(j-1)}$ similarly to (\ref{Fdeltaj}), 
is averaged with the weight
\begin{eqnarray}
\Phi^{(j)}\left(s_{1},s_{2},s_{3},\xi\right) =
e^{-if^{(j-1)}_{s}\left(s_{1},s_{2},s_{3},\xi\right)}P_{\xi}^{(j)}(\xi)
P_{s_{m}}^{(j)}(s_{1})P_{s_{m}}^{(j)}(s_{2}) P_{s_{m}}^{(j)}(s_{3}).
\end{eqnarray}
In the case if $m=1$ this yields the distributions for the nearest neighbor 
spacings. The resulting distributions can be used to compute the form factor,
$K^{(j)}_{2}(\tau)$, the autocorrelation $R^{(j)}_{2}(x)$ function and so on.

\section{Trigonometric sums and spectral universality}

An important advantage of obtaining the statistical description of spectra in 
terms of the harmonic expansions, is that the well known universality features
of the quantum chaotic systems (e.g. \cite{BGS,ABS}) can now be analyzed within 
the context of the theory of trigonometric series and weakly dependent random 
variables (see \cite{Proxorov,Revesz} and references therein). As shown e.g. in 
\cite{Erdos,SZ1,SZ2,Philipp,Philipp2,Berkes1,Fukuyama,Takahashi}, under certain 
conditions separate terms of trigonometric (or more general \cite{Berkes}) series 
statistically behave as a set of weakly dependent random variables \cite{Kahane}.
This allows establishing analytically certain universal features for the distribution 
of their sums, e.g. the corresponding generalization of the central limit theorem.

It is well known (\cite{Erdos,SZ1,SZ2,Philipp,Philipp2,Berkes1,Fukuyama,Takahashi}),
that if the frequencies of the trigonometric series
\begin{equation}
f(x)=\sum_{k}c_{k}\cos(2\pi n_{k}x+\phi_{k})
\label{trigser}
\end{equation}
increase sufficiently rapidly so that the ``critical condition'',
\begin{equation} 
\frac{n_{k+1}}{n_{k}}=1+\frac{\alpha_{k}}{\sqrt{k}}, \ \ \alpha_{k}\rightarrow\infty
\label{lacunary}
\end{equation}
is satisfied, then a central limit theorem holds for the sum (\ref{trigser}),
\begin{equation}
\frac{1}{\Delta_{K}}\sum_{k=1}^{K}c_{k}\cos(2\pi n_{k}x+\phi_{k})\rightarrow{\cal{N}}_{0,1},
\label{Gaussian}
\end{equation} 
where ${\cal{N}}_{0,1}$ is the standard normal distribution and 
$\Delta^{2}_{K}=\frac{1}{2}\sum_{k=1}^{K}c^{2}_{k}$. 

Although the series (\ref{trigser}) is much simpler than the spectral expansions
studied in the context of the periodic orbit theory, similar result was previously 
observed in the physical literature. Based on the extensive numerical simulations, 
it was hypothesized in \cite{ABS} that the fluctuations of the spectral staircase 
$\delta N(k)$ of a{\em generic} quantum chaotic system are normally distributed 
with the standard deviation
\begin{equation}
\Delta_{\infty}=\frac{1}{2}\sum_{p}A^{2}_{p},
\label{deviation}
\end{equation}
introduced in \cite{ABS} via spectral rigidity, where $A_{p}$'s are the expansion 
coefficients (\ref{series}) for $\delta N(k)$. In particular, this applies to the 
quantum graphs, for which there exists an exact periodic orbit expansion (\ref{series}) 
for $\delta N(k)$. Although the fluctuations of the quantum graph spectrum are finite 
\cite{Anima}, their distributions are typically well approximated by the Gaussian 
distribution (see Fig.\ref{Fs_Quadrangle} and Fig.\ref{Ksi_Gauss} below), just as 
the finite range $P_s$ can approximated by the Wignerian distribution \cite{QGT2} 
(see Fig.\ref{Separations3}).

The explicit expansions (\ref{zx0}) and (\ref{zxj}) allow to extend the scope of 
the ``central limit theorem for spectral fluctuations'' of \cite{ABS} hypothesis 
to a much wider range of spectral characteristics. For example, the probability 
distributions of the nearest neighbor average, $\xi_{n}^{(j-1)}$, shown in Fig.
\ref{Ksi_Gauss}, which can be obtained from (\ref{ksij}) as
\begin{eqnarray}
P_{\xi}^{(j-1)}(\xi )=\int dk\,e^{ik\xi}\left\langle F_{\xi}^{(j)}
\right\rangle_{\Upsilon_{\xi}^{(j)}}
\end{eqnarray}
with the weight
\begin{eqnarray}
\Upsilon_{\xi}^{(j)}\left(s_{1},s_{2},\xi\right) =e^{-ikf^{(j-1)}_{\xi}
\left(s_{1},s_{2},\xi\right)}P_{\xi}^{(j)}(\xi )
P_{s}^{(j)}(s_{1})P_{s}^{(j)}(s_{2}),
\end{eqnarray}
are also well approximated by Gaussian distribution with the corresponding variance.
%%%%%%%%%%%%%%%%%%%%%%%%%%%%%%%%%%%%%%%%%%%%%%%%%%%%%%%%%%%%%%%%%%%%%%
\begin{figure}[tbp]
\begin{center}
\includegraphics{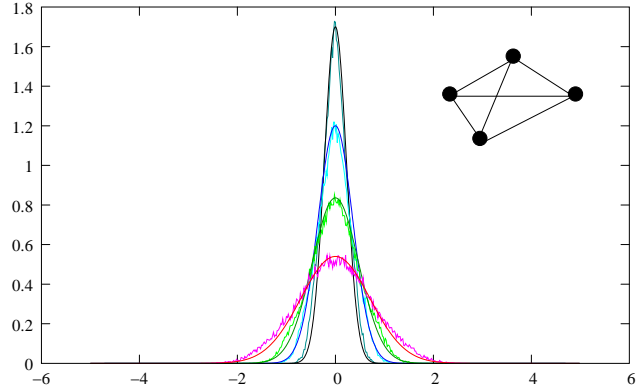}
\end{center}
\caption{Histogram of the nearest neighbor average for the even levels 
of the hierarchy of the fully connected quadrangle graph of the irregularity 
degree 7, obtained for 80,000 roots of the corresponding spectral equations. 
The solid line in the background of each histogram represents a Gaussian fit
with $\Delta$ given by the corresponding sum (\ref{deviation}).}
\label{Ksi_Gauss}
\end{figure}
%%%%%%%%%%%%%%%%%%%%%%%%%%%%%%%%%%%%%%%%%%%%%%%%%%%%%%%%%%%%%%%%%%%%%%

It is important to mention that although the condition (\ref{lacunary}) is 
required for the general proof of the central limit theorem (\ref{Gaussian}), 
\cite{Erdos,SZ1,SZ2,Philipp,Fukuyama,Takahashi}, in some cases it is possible to 
establish the existence of limit distributions if (\ref{lacunary}) is violated 
(see e.g. \cite{Berkes1,Berkes2,Levizov,Wang}). This is essential for the periodic 
orbit expansions, because lacunarity condition (\ref{lacunary}) generally may not 
hold for the prime periodic orbit spectrum of the quantum graphs and of more general 
systems. However extensive numerical and empirical evidence points on the existence 
of the corresponding limiting distributions \cite{ABS}.

There are much fewer proved results about the probabilistic behavior of the multiple 
trigonometric series such as the spectral expansions (\ref{zx0}) or (\ref{zxj}) (see 
\cite{Gaposhkin1,Gaposhkin2,BerPhil,Ulyanov,Golubov,Zygmund,Wang} and the references 
therein), specifically regarding the conditions required for convergence of their sums 
to the limiting distributions. Nevertheless, the proposed connection to the spectral 
theory clearly implies the existence of the limiting distributions for such expansions 
and points out the origins of the spectral universality \cite{BGS,ABS} from the 
perspective of the periodic orbit theory.

The nontrivial role of the transitions between the probability distributions at 
different levels of the hierarchy (such as (\ref{avedeltaj}) or (\ref{avesnmj})) 
can be seen on the example of the development of the nearest neighbor distributions
(see Fig.\ref{Separations3}). Although the distribution at the regular level has 
overall Gaussian shape, the profile of the distributions $P^{(j)}_{s}(s)$ for $j>0$ 
becomes progressively closer to the Wignerian form.

\section{Discussion}

The method of obtaining the explicit semiclassical expansions for the individual 
spectral points outlined above is based on the possibility to tie the sequence of 
momentum eigenvalues, $k_{n} \equiv \hat k_{n}^{(0)}$, to a certain base, regular 
sequence $\hat k_{n}^{(r+1)}$ defined explicitly as a function of the index $n$. 
As shown in \cite{Anima}, this can be done either directly, as in the case 
of the regular graphs, where the periodic sequence (\ref{base}) is used to place 
the points $k_{n}$ into a system of periodic cells, or indirectly, as in the case 
of the irregular graphs, where a few auxiliary sequences, $\hat k_{n}^{(j)}$, are 
needed to complete the bootstrapping.

The system of the auxiliary sequences, $\hat k_{n}^{(j)}$, that links the base 
sequence $\hat k_{n}^{(r+1)}$ and the physical spectral sequence $\hat k_{n}^{(0)}$, 
together with the rule of transition from the $j$th to $j-1$th level, defines the 
spectral hierarchy. The depth of the hierarchy, i.e. the minimal number of the 
auxiliary sequences necessary to complete the bootstrapping, expresses the complexity 
of the spectral problem with respect to a particular bootstrapping method.

The spectral hierarchy used in \cite{Anima} and in this paper is based on using 
the sequences of the roots of the derivatives of the spectral determinant and the 
base sequence 
\begin{eqnarray}
\hat k^{(r+1)}_{n}=\frac{\pi}{L_{0}}\left(n+\frac{1}{2}\right).
\label{base}
\end{eqnarray}
The index $n\in \mathbb{N}$ that appears explicitly in (\ref{base}) is then carried 
to the $0$th level via the rule (\ref{seps}). This scheme allows describe the exact 
evolution of the separating sequences $k_{n}^{(j)}$ from the lower to the upper levels 
of the hierarchy, and to pass on to a general probabilistic description of the spectral 
characteristics, including the ones that not accessible via the Gutzwiller's expansion 
for the density of states.

The organization of the spectral hierarchy allows one to follow the accumulation of the 
fluctuations from the regular ($r$th) to the physical ($0$th) level. Essentially, this 
method allows to unfold the full scale spectral fluctuations in steps, by distributing 
the disorder across the intermediate levels of the hierarchy, and by passing successively 
from more orderly to more disordered sequences. While the starting sequence is perfectly 
ordered (e.g. (\ref{base}) is periodic), every other sequence $\hat{k}_{n}^{(j)}$, $j<r+1$, 
is disordered, and the scale of the oscillations increases as one passes from the level $j$ 
to $j-1$ \cite{Anima}. The probability distributions at every level are obtained by averaging 
over the fluctuations introduced by the periodic orbits at that level, as well as over the 
disorder $\delta^{(j)}$ inherited from the previous level of the hierarchy.

The spectral expansions may also provide a physical understanding of the origins of the 
spectral statistical universalities based on the periodic orbit theory. It is well known 
that under certain conditions (e.g. the lacunarity condition (\ref{lacunary})), separate 
terms or specially combined groups of terms of the trigonometric series behave as weakly 
dependent random variables (\cite{Proxorov,Revesz}). This allows to establish standard 
universal asymptotic distributions for their sums, in particular convergence to a Gaussian 
distribution with a certain specific variance. Interestingly, the same variance was already 
conjectured for the universal probability distribution profile for the $\delta^{(0)}_{n}$ 
fluctuations of a generic quantum chaotic systems \cite{ABS}).

In addition, in the proposed approach, the propagation of the fluctuations through the hierarchy 
and the build-up of the distributions $P^{(j)}$ via the corresponding number of averagings over 
the disordered sequences $\hat k_{n}^{(j)}$ can lead to the appearance of other universal (e.g. 
Wignerian) profiles, as it is the case for the nearest neighbor separation statistics (see Fig.
\ref{Separations3}).

On the other hand, it is also important that this approach does not overlook the individual 
features of a particular system for the sake of broad universality, and can provide detailed
description of the distributions that deviate from the universal behavior (as in the case of
the regular graphs), as well as the degree of such deviation.

%%%%%%%%%%%%%%%%%%%%%%%%%%%%%%%%%%%%%%%%%%%%%%%%%%%%%%%%%%%%%%%%%%%%%%%%%
The work was supported in part by the Sloan and Swartz Foundation.
%%%%%%%%%%%%%%%%%%%%%%%%%%%%%%%%%%%%%%%%%%%%%%%%%%%%%%%%%%%%%%%%%%%%%%%%%

\end{document}